\documentclass[showpacs,preprintnumbers,amsmath,amssymb,aps, prd]{revtex4}
\usepackage{graphicx}
\usepackage{grffile}
\usepackage{mathrsfs, mathtools}
\usepackage{caption}
\usepackage{float}
\usepackage{xcolor}
\usepackage{changepage}
\usepackage{dcolumn}
\usepackage{bm}
\usepackage{graphicx,subfigure,epsfig}
\usepackage{multirow}

\def\mn{_{\mu\nu}}

\def\a{\alpha}

\def\f{\frac}

\def\e{\mathcal{E}}
\def\l{\mathcal{L}}
\def\grr{g_{rr}}
\def\gtt{g_{tt}}
\def\gth{g_{\phi\phi}}
\def\veff{\mc{V}_{\text{eff}}}

\def\ab{\overline{a}}

\def\bb{\overline{b}}
\def\tq{\tilde{Q}}
\def\c{\cite}
\def\r{\ref}
\def\fr{F(r)}
\def\hr{H(r)}
\def\ar{A(r)}

\def\fx{F(x)}
\def\hx{H(x)}
\def\s{Schwarzschild }

\newcommand\be{\begin{equation}}
\newcommand\ee{\end{equation}}
\newcommand\ba{\begin{eqnarray}}
\newcommand\ea{\end{eqnarray}}
\newcommand\bt{\bibitem}
\newcommand\nn{\nonumber}
\newcommand\lt{\left}
\newcommand\rt{\right}
\newcommand\pt{\partial}
\newcommand\tx{\text}
\newcommand\mc{\mathcal}
\begin{document}
\title{Strong gravitational lensing and Quasiperiodic oscillations as a probe for an electrically charged Lorentz symmetry-violating black hole}
\author{Sohan Kumar Jha}
\email{sohan00slg@gmail.com}
\affiliation{Department of Physics, APC Roy Govt. College, Siliguri, West
Bengal, India}

\date{\today}
\begin{abstract}
\begin{center}
Abstract
\end{center}
This study examines the combined effect of electric charge and Lorentz symmetry breaking (LSB) on the observables of strong gravitational lensing (SGL) and the dynamics of quasiperiodic oscillations (QPOs) around an electrically charged, Lorentz symmetry-violating (LV) black hole (QKR BH). We first explore the SGL, which unravels an interesting effect that the two combined generate. We find cases where the competing effect of charge and LV cancels each other, leaving the underlying quantity unchanged from that of a \s BH. We find bounds on the LV parameter utilizing observations related to the shadow angular size of supermassive black holes (SMBHs) $M87^*$ and $SgrA^*$. No bound could be gleaned for the charge from these shadow observations. Observations of QPOs in microquasars provide an alternative method to probe our model and to extract bounds on its parameters. We use experimental data for the microquasars $GRO J1655-40$ and $XTE J1550-564$. Here we obtain bounds on both parameters. Our results provide deeper insights into the interplay between charge and LSB in the strong-gravity regime.  
\end{abstract}
\maketitle
\section{Introduction}
Despite the astounding success of general relativity (GR) and results of numerous experiments confirming non-violation of Lorentz symmetry, numerous theoretical models motivated by string theory, non-commutative field theory, and others predict Lorentz symmetry's violation at the fundamental scale [\citenum{Kostelecky1989a}-\citenum{Cohen2006}]. In the case of explicit LSB, the Lagrangian density no longer remains invariant under Lorentz transformation, resulting in non-invariability of physical laws in different inertial frames. On the other hand, in spontaneous LSB, physical laws remain invariant in all inertial frames, but the Lorentz symmetry is broken in the ground state. The LSB effect is elegantly incorporated in the standard model extension (SME) \cite{Kostelecky2004a}. Within SME, LSB is induced through a non-zero vacuum expectation value (VEV) of either a vector \cite{Kostelecky1989a, Kostelecky1989, Kostelecky1989b, Bailey2006, Bluhm2008a} or an antisymmetric tensor of rank two \c{action1}. Authors in \c{bm} obtained an exact solution for the former, known as the bumblebee model, whereas the latter was adopted by authors in \c{action2} and \c{kr}. Please see [\citenum{Ovgun2018}-\citenum{Maluf2022a}] for various studies pertaining to LV BHs. This study considers an electrically charged LV BH reported in \c{qkr} which, in the absence of the LV effect, resorts to the metric in \c{kr}.\\
Strong gravitational lensing has been extensively employed to probe the strong-field regime near a black hole, as it provides valuable information about the intrinsic properties of the background spacetime. Since modified theories of gravity must conform with GR when the weak field limit is taken, studies related to strong fields provide avenues to gauge divergence from GR. This makes SGL such a potent tool for testing modified theories and their divergence from GR in the strong-field limit. When a photon passes by a compact object, such as a black hole, its path is deflected by the black hole's strong gravitational pull. This deflection of a photon's trajectory is known as gravitational lensing (GL), and the compact object is the gravitational lens. Darwin pioneered the application of GL to BH in his seminal work \c{dar} where he considered the \s BH. Studies such as \c{vir} and \c{BOZZA} shed more light and add more tools to our arsenal for studying SGL. Numerous studies have used SGL to assess the deviation of the models under consideration from GR. Please refer [\citenum{BOZZA1}-\citenum{ikr1}] for SGL's application to numerous spacetimes. Experimental observations provide a unique opportunity to extract bounds on free parameters. This study utilizes angular diameter-related observations of SMBHs $M87^*$ and $SgrA^*$ to test the commensurability of our model and glean constraints on model parameters \c{akiyamal1, akiyamal12, sgra}.\\
Another excellent avenue for testing gravity near BHs is via QPOs. They frequently appear as narrow peaks in the X-ray power density spectra of binary systems such as BHs and neutron stars (NSs). Since these oscillations occur in the vicinity of compact objects, analyzing them would provide valuable information about spacetime geometry and could thus be utilized to test alternative theories of gravity. Within these oscillations, there exist high-frequency (HF) QPOs with a typical range of $50-1300$ Hz. Since this range is of the order similar to the frequencies related to orbital motion in proximity to compact objects, most models intending to explain HF QPOs take orbital motions in the accretion disc's innermost regions into consideration. HF QPOs commonly occur as twin peaks with upper $\nu_{U}$ and lower $\nu_L$ frequencies exhibit $3:2$ ratio \c{qpo40, qpo401, qpo564}. They have been observed in the X-ray flux of numerous microquasars such as $GRO J1655-40$ and $XTE J1550-564$. Several theoretical models, such as Keplerian resonance, parametric resonance, and forced resonance, have been proposed to explain observed HF QPOs \c{resonance, resonance1, forced}. We employ the forced resonance model in this study. Our main concern in this article is to extract bounds on free parameters using HF QPOs of microquasars $GRO J1655-40$ and $XTE J1550-564$. Please refer [\citenum{sanjar}-\citenum{QPO13}] for more details on the application of QPO in various spacetimes.\\
We structure this study as follows. Sec. II is where we briefly introduce the metric in concern and obtain conditions for its existence. In Sec. III and IV, we deal with strong gravitational lensing in the background of a QKR BH and try to constrain free parameters using SGL observables. Sec. V is where we study the properties of QPOs and use observations of HF QPOs to bound model parameters. We conclude our article in Sec. VI with a brief overview of the obtained results and future prospects. We have used $G=c=M=1$ throughout the paper.
\section{electrically charged lorentz violating bh}
Here we briefly introduce the electrically charged Lorentz-violating static and spherically symmetric BH solution obtained in \c{qkr}. The non-zero VEV of the KR field $B\mn$ is responsible for the spontaneous breaking of Lorentz symmetry in this case. The action for the considered model is \c{qkr}:
\begin{widetext}
\begin{align}\label{action}
S=\int d^4x\sqrt{-g}\bigg[\frac{1}{2}\bigg(R-2\Lambda+\varepsilon\, B^{\mu\lambda}B^\nu\, _\lambda R_{\mu\nu}\bigg)-\frac{1}{12}H_{\lambda\mu\nu}H^{\lambda\mu\nu}-V(B_{\mu\nu}B^{\mu\nu}\pm b^2)+\mathcal{L}_{m}\bigg],
\end{align}
\end{widetext}
where $H^{\mu\nu\rho}$ is the KR field strength, $\varepsilon$ is the coupling constant that arises due to non-minimal coupling between the KR field and gravity, and $V(B^{\mu\nu}B_{\mu\nu}\pm b^2)$ is the self-interacting potential that ensures the Lorentz symmetry breaking through generation of non-zero VEV of the field $B\mn$. The Lagrangian density for matter $\mathcal{L}_{m}$ is:
\begin{equation}
\mathcal{L}_{m}=-\frac{1}{2}F^{\mu\nu}F_{\mu\nu}-\eta B^{\alpha\beta}B^{\gamma\rho}F_{\alpha\beta}F_{\gamma\rho},
\end{equation}
where $F_{\mu\nu}=\partial_{[\mu} A_{\nu]}$ represents the em field strength, and $\eta$ is a coupling constant that corresponds to non-minimal coupling between the em and KR fields. Field equations arising from the action (\r{action}) by varying it with respect to $g^{\mu \nu}$ are
\ba
R_{\mu \nu }-\frac{1}{2}g_{\mu \nu }R+\Lambda  g_{\mu \nu }= T^{\tx{KR}}_{\mu\nu}+T^{\tx{EM}}_{\mu\nu},
\label{fe}
\ea
where $T^{\tx{EM}}_{\mu\nu}$ is the energy-momemtum tensor for the em field given by
\ba\nn
T^{\tx{EM}}_{\mu\nu}= 2F_{\mu\alpha}F_{\nu}{}^{\alpha}-\frac{1}{2}g_{\mu\nu}F^{\alpha\beta}F_{\alpha\beta}+\eta\left(8B^{\alpha\beta}F_{\alpha\beta}B_{\left(\mu\right.}{}^{\gamma}F_{\left.\nu\right)\gamma}-g_{\mu\nu}B^{\alpha\beta}B^{\gamma\rho}F_{\alpha\beta}F_{\gamma\rho}\right),
\label{tem}
\ea
and the energy-momentum tensor for the KR field is given by
\ba
T^{\tx{KR}}_{\mu\nu} \!&\!=\!&\!\frac{1}{2} H_{\mu \alpha \beta } H_{\nu }{}^{\alpha \beta } \!-\! \frac{1}{12} g_{\mu \nu } H^{\alpha \beta \rho } H_{\alpha \beta \rho }\!+\!2V'(X) B_{\alpha\mu}B^{\alpha}{}_\nu  \nn\\
\!&\!-\!&\! g_{\mu\nu}V(X)+ \varepsilon \bigg[\frac{1}{2} g_{\mu \nu } B^{\alpha \gamma } B^{\beta }{}_{\gamma }R_{\alpha \beta } - B^{\alpha }{}_{\mu } B^{\beta }{}_{\nu }R_{\alpha \beta }\nn\\
\!&\!-\!&\!  B^{\alpha \beta } B_{\nu \beta } R_{\mu \alpha }-B^{\alpha \beta } B_{\mu \beta } R_{\nu \alpha }+\frac{1}{2} \nabla _{\alpha }\nabla _{\mu }\left(B^{\alpha \beta } B_{\nu \beta }\right)
\nn\\
\!&\!+\!&\!
\frac{1}{2} \nabla _{\alpha }\nabla _{\nu }\left(B^{\alpha \beta } B_{\mu \beta }\right)-\frac{1}{2}\nabla ^{\alpha }\nabla _{\alpha }\left(B_{\mu }{}^{\gamma }B_{\nu \gamma } \right)
\nn\\
\!&\!-\!&\! \frac{1}{2} g_{\mu \nu } \nabla _{\alpha }\nabla _{\beta }\left(B^{\alpha \gamma } B^{\beta }{}_{\gamma }\right)\bigg],
\label{tkr}
\ea
where the prime connotes the derivative of functions with respect to their respective arguments.\\
Equations of motion for the em and KR fields are, respectively, given by
\ba
\nabla^\nu\left(F_{\mu\nu}+2\eta B_{\mu\nu}B^{\alpha\beta}F_{\alpha\beta}\right)&=&0,\label{em}\\
\nabla^{\alpha}H_{\alpha\mu\nu}+3\varepsilon R_{\alpha[\mu}B^{\alpha}{}_{\nu]}-6V^{\prime}B_{\mu\nu}-12\eta B^{\alpha\beta}F_{\alpha\beta}F_{\mu\nu}&=&0\label{kr}.
\ea
Solving Eq. (\r{fe}), (\r{em}), and (\r{kr}) simultaneously provide the following metric
\be
ds^2=-\ar~dt^2+\f{dr^2}{\ar}+r^2 d\theta^2+r^2 \sin^2\theta d\phi^2,
\label{metric}
\ee
with $\ar = \f{1}{1-\a}-\f{2M}{r}+\f{Q^2}{(1-\a)^2 r^2}$. In the limit $\a \rightarrow 0$, we recover the Reissner-Nordstr{\"o}m (RN) metric, whereas taking a further limit $Q \rightarrow 0$ produces metric for the \s BH.\\
We obtain the position of the event horizon by solving the equation $\ar=0$ and retaining the largest root of the resultant equation. The subsequent expression for the event horizon is
\be
r_h= M(1-\a)+(1-\a)\sqrt{M^2 - \frac{Q^2}{(1-\a)^3}}\label{event}.
\ee
The QKR BH has an inner horizon located at
\be
r_c= M(1-\a)-(1-\a)\sqrt{M^2 - \frac{Q^2}{(1-\a)^3}}.
\ee
The horizons exist when the BH parameters satisfy the condition
\be
M^2 - \frac{Q^2}{(1-\a)^3}\ge 0 \quad \Rightarrow \quad Q^2 \le M^2 (1-\a)^3.\label{qb}
\ee
When the equality holds in the above condition, two horizons occur at the same position, and we have an extrmal QKR BH.
\begin{figure}[H]
\begin{center}
\begin{tabular}{c}
\includegraphics[width=0.4\columnwidth]{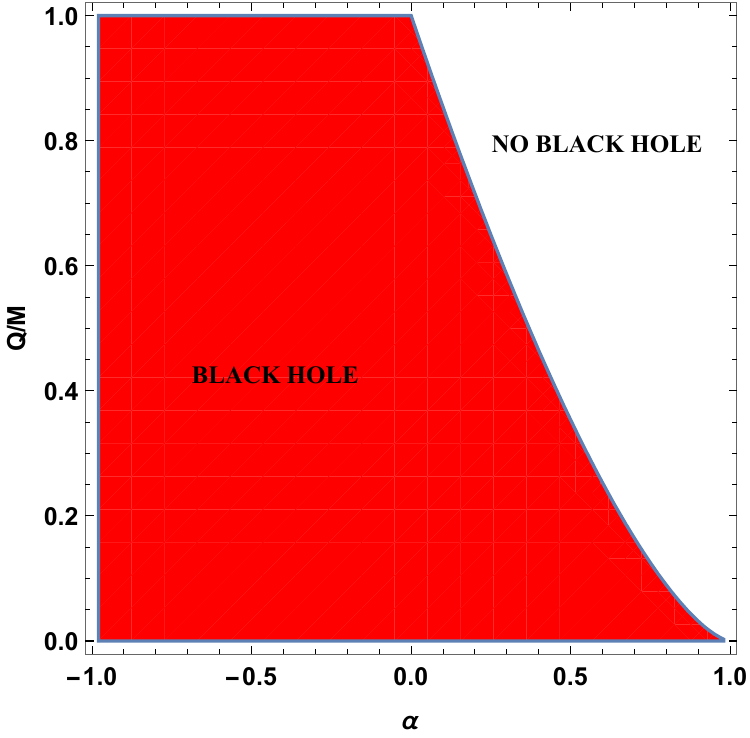}
\end{tabular}
\caption{Parameter space for the existence of BH. The coloured region is where we have a BH solution. }\label{para}
\end{center}
\end{figure}
Fig. (\r{para}) illustrates the parameter space $(\a,~Q/M)$ for which we have a BH solution. Expression for the event horizon (\r{event}) points towards the adverse impact of charge on $r_h$, whereas the impact of the LV parameter depends on its sign. For a negative value $\a$, $r_h$ for a QKR BH is always greater than that for an RN BH, and the reverse situation is true for positive values of $\a$. As such, it is possible to find combinations of $(\a, ~Q/M)$ with $\a<0$ where the incrementing effect of negative $\a$ offsets the decrementing effect of charge, thereby leaving the event horizon equal to that for a \s BH. For illustrative purposes, the impact of charge $Q$ and LV parameter $\a$ on the event horizon is displayed in Fig. (\r{rh}).
\begin{figure}[H]
\begin{center}
\begin{tabular}{cc}
\includegraphics[width=0.4\columnwidth]{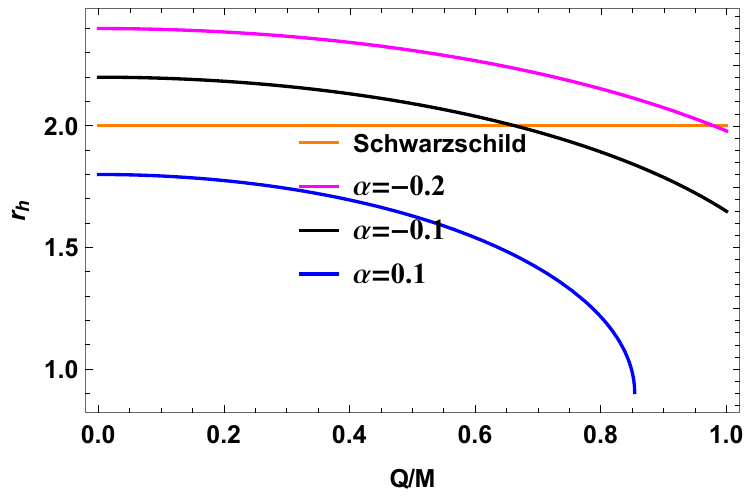}&
\includegraphics[width=0.4\columnwidth]{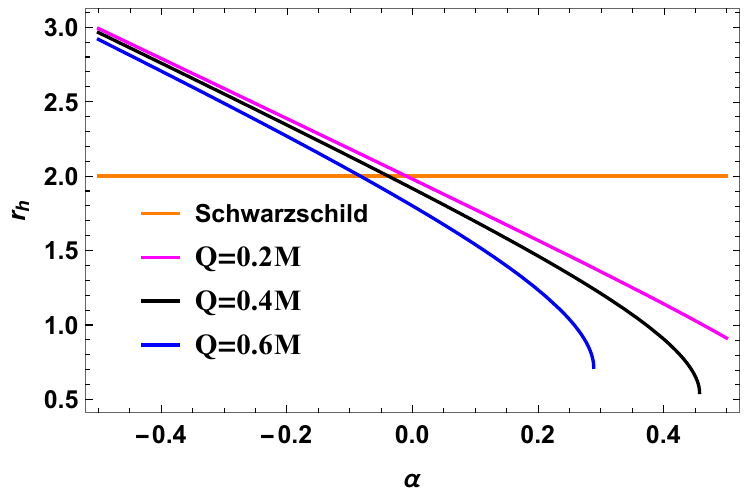}
\end{tabular}
\caption{Variation of event horizon with charge $Q$ (left panel) and with LV parameter $\a$ (right panel). }\label{rh}
\end{center}
\end{figure}
Both parameters $Q$ and $\a$ adversely impact $r_h$. Fig. (\r{rh}) demonstrates the existence of combinations $(\a, ~ Q/M)$ where the event horizon for a QKR BH occurs at the same position as that of a \s BH. Few such combinations are $(-0.1,~0.66M)$, $(-0.04,~0.4M)$, and $(-0.0099,~0.2M)$. Having introduced the metric to be probed, we move on to our first theme of this paper: strong gravitational lensing and its application in constraining parameters $Q$ and $\a$.
\section{strong gravitational lensing in the background of a qkr bh}
The strong gravitational field of a BH deflects even photons from their original course. Such gravitationally lensed photons carry imprints of the intrinsic properties of the underlying spacetime and encode valuable information about the massive central object. This makes SGL such a potent tool to test any proposed model. We follow the prescription expounded in \cite{BOZZA, BOZZA1, BOZZA2}. We concern ourselves only with the equatorial plane and rewrite the metric (\r{metric}) as
\begin{equation}
d\tilde{s}^{2}=(2M)^{-2}ds^{2}=-F(x) dt^{2}+F(x)^{-1} dx^{2}+H(x) d \phi^{2}, \label{final1}
\end{equation}
where $x=\f{r}{2M}$, $\tq=\f{Q}{2M}$, and
\be
\fx= 1-\f{1}{x}+\f{\tq^2}{(1-\a)^2 x^2}\quad \tx{and} \quad \hx=x^2.
\ee
The Lagrangian associated with the metric (\r{final1}) is
\be
\l=\f{1}{2}\lt(-\fx \dot{t}^2+\f{\dot{x}^2}{\fx}+\hx \dot{\phi}^2\rt),\label{l}
\ee
where $\dot{t}$ and $\dot{\phi}$ are, respectively, differentiation of $t$ and $\phi$ with respect to the affine parameter $\Lambda$. As the Lagrangian does not explicitly depend on $t$ and $\phi$, two quantities - energy $E$ and angular momentum $L$ are conserved along null geodesics. They are given by
\be
E=-\f{d\l}{d\dot{t}}=\fx \dot{t} \quad \tx{and} \quad L=\f{d\l}{d\dot{\phi}}=\hx \dot{\phi}.
\ee
The above expression, conjoined with $d\tilde{s}=0$ for null geodesics, eventually leads us to the following equation:
\be
\dot{x}^2=E^2-\f{L^2\fx}{\hx}=E^2-V(x),
\ee
where the potential $V(x)=\f{L^2\fx}{\hx}$ governs the photon trajectory. The condition $\f{dV}{dx}|_{x=x_m}=0$ for a circular orbit of radius $x_m$ leads us to the expression for the photon orbits' radius as:
\be
H'(x_m)F(x_m)=F'(x_m)H(x_m) \quad \rightarrow \quad x_m=\frac{\sqrt{\alpha -1} \sqrt{32 \tilde{Q}^2+9 \alpha ^3-27 \alpha ^2+27 \alpha -9}-3 \alpha ^2+6 \alpha -3}{4 (\alpha -1)}.
\ee
where prime represents $\f{d}{dx}$. Photons can only approach the BH up to a minimum distance $x_0$, depending on its impact parameter $b$, following the relation:
\be
\f{dx}{d\phi}=0 \quad \Rightarrow \quad b=\f{L}{E}=\sqrt{\f{H(x_0)}{F(x_0)}}.
\ee
The minimum value of $x_0$ for those photons which contribute to the optical appearance of the BH is $x_m$, and the associated impact parameter is $b_m$. Only those photons having $b>b_m$ get deflected towards the asymptotic observer, whereas those with $b<b_m$ can not escape the strong gravitational pull of the BH. Photons with impact parameter very close to $b_m$ are responsible for the bright ring that marks the outline of the BH shadow. As such, an asymptotic observer sees a BH shadow of radius $b_m$. In the limit $\a \rightarrow 0$ and $\tq \rightarrow 0$, we obtain $b_m=\f{3\sqrt{3}}{2}$ which is the shadow radius for a \s BH. In fig. (\r{bm}) we explore the variation of shadow radius with the LV parameter $\a$ and charge $\tq$.
\begin{figure}[H]
\begin{center}
\begin{tabular}{cc}
\includegraphics[width=0.4\columnwidth]{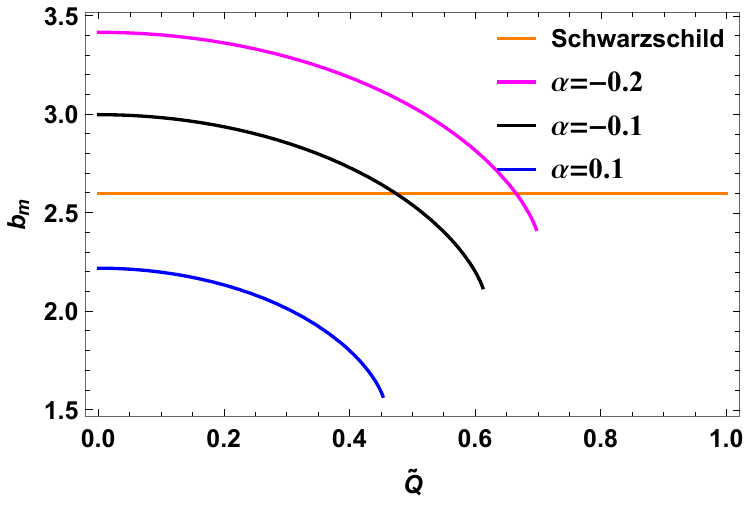}&
\includegraphics[width=0.4\columnwidth]{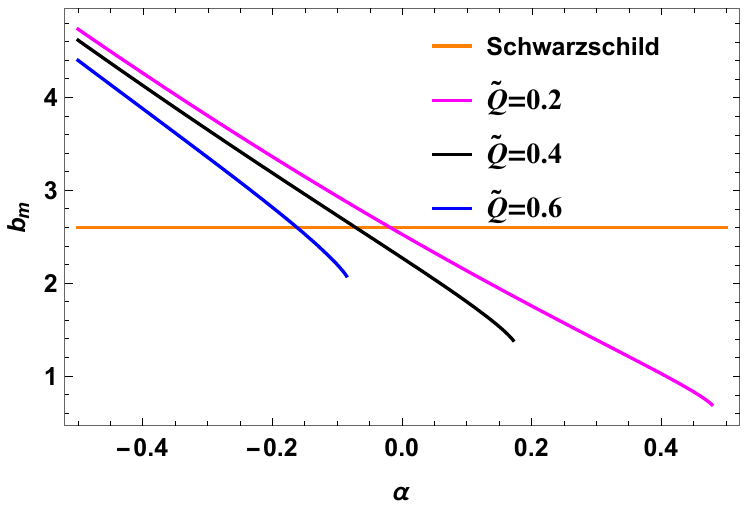}
\end{tabular}
\caption{Variation of shadow radius with charge $\tq$ (left panel) and with LV parameter $\a$ (right panel). }\label{bm}
\end{center}
\end{figure}
Similar to the event horizon case, a QKR BH with a negative LV parameter has a larger shadow radius than an RN BH, whereas in comparison to an RN BH, a QKR BH's shadow radius may be greater, equal, or less, depending on the charge value. For example, at $\a=-0.1$, the shadow radii of QKR and \s BHs become equal when $\tq=0.473451$. For $\tq$ less than this, we have a QKR BH whose shadow radius is larger than a \s BH, whereas the situation is reverse for charge greater than $0.473451$. Therefore, a transition is seen in this regard. When $\a>0$, a QKR BH's shadow is always smaller than a \s BH. \\
The deflection angle, following articles \c{vir, wein}, can be written as
\begin{equation}
\alpha_D\left(x_{0}\right)=I\left(x_{0}\right)-\pi,\label{def}
\end{equation}
where
\begin{equation}
I\left(x_{0}\right)=\int_{x_{0}}^{\infty}\frac{2}{\sqrt{F(x)H(x)}
\sqrt{\frac{F(x_{0})H(x)}{H(x_{0})F(x)}-1}}dx.\label{io}
\end{equation}
Since the integral (\r{io}) diverges at $x_0=x_m$, a new variable $z=1-\f{x_0}{x}$ is considered. Taking cue from \c{BOZZA} leads us to the deflection angle as:
\begin{eqnarray}
&& \gamma_D(b)=-\overline{a} \log \left( \frac{b}{b_m} -1
\right) +\overline{b}+O(b-b_m) ,\label{alphab}\\\nn%
\text{where}\\
&& \overline{a}=\frac{a}{2}=\frac{\mathcal{R}(0,x_m)}{2\sqrt{a_2(x_m)}}, \label{afinal}\\%
&& \overline{b}=-\pi+\bar{b}_\mathcal{R}+\overline{a}\log{\frac{2H^2(x_m)a_2(x_m)}{F(x_m)x_m^4}}\label{bfinal},\\\nn
\text{with}\\\nn
&& \mathcal{R}(z,x_m)=\frac{2x^{2}\sqrt{H(x_0)}}{x_{0}H(x)},\\\nn
&&a_{2}(x_0) = \frac{1}{2}\left[\frac{\lt(2x_{0}H(x_{0})-2x_{0}^{2}H^{\prime}(x_{0})\rt)\lt(H^\prime(x_0) F(x_0)-F^\prime(x_0) H(x_0)\rt)}{H^{2}(x_{0})}+\frac{x_{0}}{H(x_{0})}\left(H^{\prime\prime}(x_{0})F(x_{0})-F^{\prime\prime}(x_{0})F(x_{0})\right)\right],\\\nn
&&g(z,x_0)=\mathcal{R}(z,x_0)f(z,x_0)-\mathcal{R}(0,x_m)f_0(z,x_0),\\\nn
&& I_\mathcal{R}(x_0)=\int\limits_0^1 g(z,x_m) dz+O(x_0-x_m) \quad \text{and} \quad \bar{b}_\mathcal{R}= I_\mathcal{R}(x_m).
\end{eqnarray}
Lensing coefficients given by eq. (\r{afinal}) and (\r{bfinal}), in the absence of LV and charge, take the values $\ab=1$ and $\bb=-0.40023$, their values for the \s BH reported in \c{BOZZA}. Table (\r{abv}) displays how the variation of model parameters $\a$ and $\tq$ affects lensing coefficients and shadow radius. Here, $\delta$ of any quantity denotes its deviation from the \s case. It can easily be construed from the table that while for zero or positive values of the LV parameter, these quantities are smaller than those for the \s BH, they are larger when the LV parameter takes a negative value.
\begin{table}[H]
\begin{centering}
\begin{tabular*}{\textwidth}{@{\extracolsep{\fill}\quad}cccccccc}
\hline\hline
{$\alpha$ }& {$\tq$} & {$\ab$} & {$\delta \ab$} & {$\bb$} & {$\delta \bb$} & {$b_m$} & {$\delta b_m$}
\\ \hline
\hline
\multirow{5}{*}{-0.1}&$0$ & $1.04881$ & $0.0488088$ & $-0.266427$ & $0.133803$ & $2.99737$ & $0.399298$ \\
&$0.1$ & $1.05238$ & $0.0523767$ & $-0.265726$ & $0.134504$ & $2.98227$ & $0.384196$ \\
&$0.2$ & $1.06394$ & $0.0639413$ & $-0.263884$ & $0.136346$ & $2.93585$ & $0.33777$ \\
&$0.3$ & $1.08677$ & $0.0867695$ & $-0.262179$ & $0.138051$ & $2.85426$ & $0.25618$ \\
&$0.4$ & $1.12998$ & $0.129976$ & $-0.26589$ & $0.13434$ & $2.72892$ & $0.130843$ \\
\hline
\multirow{5}{*}{0}&$0$ & $1.$ & $0.$ & $-0.40023$ & $0$ & $2.59808$ & $0$ \\
&$0.1$ & $1.00456$ & $0.00455621$ & $-0.399348$ & $0.000882238$ & $2.58062$ & $-0.0174574$ \\
&$0.2$ & $1.01974$ & $0.0197361$ & $-0.397184$ & $0.00304588$ & $2.52649$ & $-0.0715876$ \\
&$0.3$ & $1.05183$ & $0.0518265$ & $-0.396509$ & $0.00372123$ & $2.42935$ & $-0.168729$ \\
&$0.4$ & $1.12317$ & $0.123169$ & $-0.413638$ & $-0.0134082$ & $2.27299$ & $-0.325083$ \\
\hline
\multirow{5}{*}{0.1}&$0$ & $0.948683$ & $-0.0513167$ & $-0.540908$ & $-0.140678$ & $2.21828$ & $-0.3798$ \\
&$0.1$ & $0.954669$ & $-0.0453313$ & $-0.539776$ & $-0.139546$ & $2.19777$ & $-0.400307$ \\
&$0.2$ & $0.975489$ & $-0.0245108$ & $-0.53736$ & $-0.13713$ & $2.13333$ & $-0.464749$ \\
&$0.3$ & $1.02508$ & $0.025075$ & $-0.54101$ & $-0.14078$ & $2.01349$ & $-0.584588$ \\
&$0.4$ & $1.18752$ & $0.187522$ & $-0.642253$ & $-0.242023$ & $1.79984$ & $-0.798234$ \\
\hline\hline
\end{tabular*}
\end{centering}
\caption{Values of lensing coefficients for different values of $\a$ and $\tq$. \label{abv}}
\end{table}
The variation of deflection angle with the impact parameter for different values of $\a$ and $\tq$ is illustrated graphically in fig. (\r{def}). Here too, we see the deflection angle is smaller than for a \s BH when $\a>0$, whereas the situation is reversed for a negative LV parameter. The charge displays an adverse impact on the deflection angle.
\begin{figure}[H]
\begin{center}
\begin{tabular}{cc}
\includegraphics[width=0.4\columnwidth]{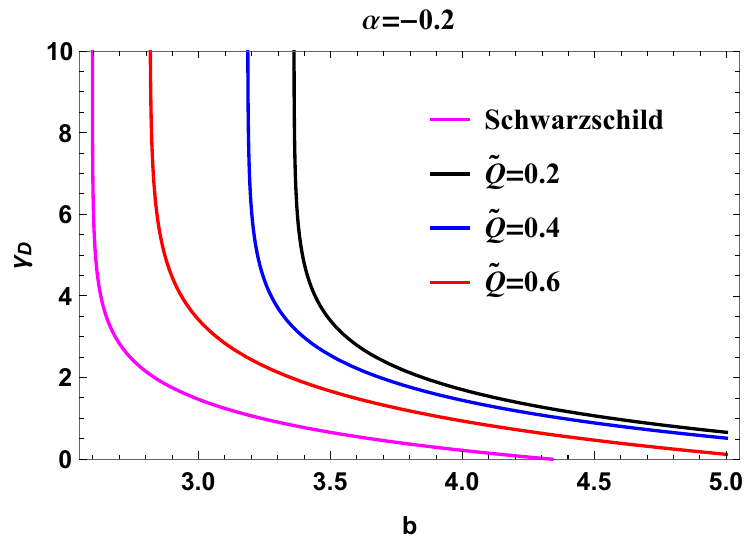}&
\includegraphics[width=0.4\columnwidth]{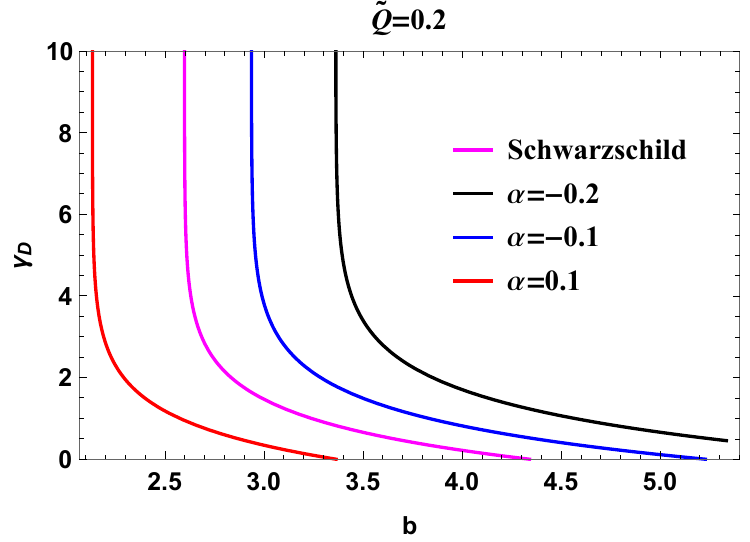}
\end{tabular}
\caption{Variation of deflection angle with impact parameter $b$ for various values of charge $\tq$ (left panel) and LV parameter $\a$ (right panel). }\label{def}
\end{center}
\end{figure}
\section{observables in strong gravitational lensing}
This section deals with observables related to SGL. We take cues from articles \c{BOZZA, BOZZA1} to this end. The lens equation governing the Lens position (here BH), positions of the source S, and the observer O is \c{BOZZA1}:
\begin{equation}
\eta=\theta-\frac{D_{LS}}{D_{OS}} \Delta \gamma_n,%
\label{lens}
\end{equation}
where $D_{LS}$ and $D_{OS}$ are, respectively, the distances of the Lens and the observer from the source, $\Delta \gamma_n=\gamma(\theta)-2n\pi$ represents the offset of the deflection angle, and $n$ is an integer that signifies number of turns a photon takes around BH before it escapes towards the observer. The angular position taken by the $n^{th}$ relativistic image is \c{BOZZA}
\begin{equation}
\theta_n = \theta_n^0 +\frac{ b_m e_n\left(\eta-\theta_n^0\right)
D_{OS}}{\overline{a} D_{LS}D_{OL}},
\label{images}
\end{equation}
where the solution of $\gamma(\theta)=2n\pi$ provides $\theta_n^0$ as
\begin{eqnarray}
\theta_n^0=\frac{b_m}{D_{OL}} \left(1+e_n \right) \label{theta0}\quad \text{where} \quad e_n=e^{\frac{\overline{b}-2n\pi}{\overline{a}}}.\label{en}
\end{eqnarray}
When $\theta_n^0$ becomes equal to $\eta$, coincidence of the image and the source occurs, and hence there exists no correction to the position of the $n^{th}$ image. Images form on the same side of the source when we consider eq. (\r{images}). Replacing $\eta$ with $-\eta$ will result in a situation where images form on the opposite side of the source. The magnification of the image provides significant information, which is defined by \c{BOZZA}
\begin{eqnarray}
\mu_n &=& \left(\frac{\eta}{\theta} \;
\;\frac{d\eta}{d\theta} \Bigg|_{\theta_n ^0}\right)^{-1}\\
&=&e_n \frac{ b_m^2\left(1+e_n \right)
D_{OS}}{\overline{a} \eta D_{OL}^2 D_{LS}}.\label{magnification}
\end{eqnarray}
Since $e_n$ decreases with the winding number $n$, so does the magnification. This results in images formed from photons with higher $n$ being faint unless $\eta\rightarrow 0$, which signifies a perfect alignment between S and L. The brightest image, which is the outermost image with angular position $\theta_1$, is considered resolved, whereas the rest are considered to be clubbed at $\theta_\infty$. This situation raises three observables:
\begin{eqnarray}
&&\theta_\infty = \frac{b_m}{D_{OL}},\quad s= \theta _1-\theta _\infty = \theta_\infty \;\; e^{\frac{\bar{b}-2\pi}{\bar{a}}},\quad \text{and \quad }r_{\text{mag}}= 2.5 \log(r) = \frac{5\pi}{\bar{a}\ln 10}\label{s}\\\nn
\text{where}\\
&&r = \frac{\mu_1}{\sum{_{n=2}^\infty}\mu_n } = e^{\frac{2 \pi}{\bar{a}}}.\label{obs}
\end{eqnarray}
Here, the angular separation between the outermost and the rest images is $s$, $r_{mag}$ connotes the magnification of the brightest image relative to the rest, and $r$ denotes the ratio of the flux from the brightest image to the flux from the rest. These observables play a significant role, as we can extract lensing coefficients $\ab$ and $\bb$ from them once we have their values from astronomical observations. This will help us decode the intricate nature of the background spacetime, which is imprinted in lensing coefficients. We intend to put our model to test against experimental findings of SMBHs $M87^*$ and $SgrA^*$ whose mass $(M)$, distance from the Earth $D$, and angular diameter $\theta_d\,(=2\theta_\infty)$ are \c{akiyamal1, akiyamal12, sgra}:
\ba\nn
&&M87^*: \quad M=(6.5\,\pm\,0.7)\,\times 10^9\,M_{\odot},\quad D=(16.8\,\pm\,0.8)\, Mpc, \quad \tx{and}\quad \theta_{d}=42\,\pm\,3 \mu as,\\\nn
&&SgrA^*: \quad M=4.28^{\pm 0.10}_{\pm 0.21}\,\times 10^6\,M_{\odot},\quad D=8.32^{\pm 0.07}_{\pm 0.14}\, kpc, \quad \tx{and}\quad \theta_{d}=48.7\,\pm\,7 \mu as.\label{data}
\ea
Here $M_{\odot}$ denotes the Sun's mass. In Fig. (\r{rmag}), we assess the effects of $\a$ and $\tq$ on the relative magnification when modeling the SMBH $SgrA^*$ as a QKR BH.
\begin{figure}[H]
\begin{center}
\begin{tabular}{cc}
\includegraphics[width=0.4\columnwidth]{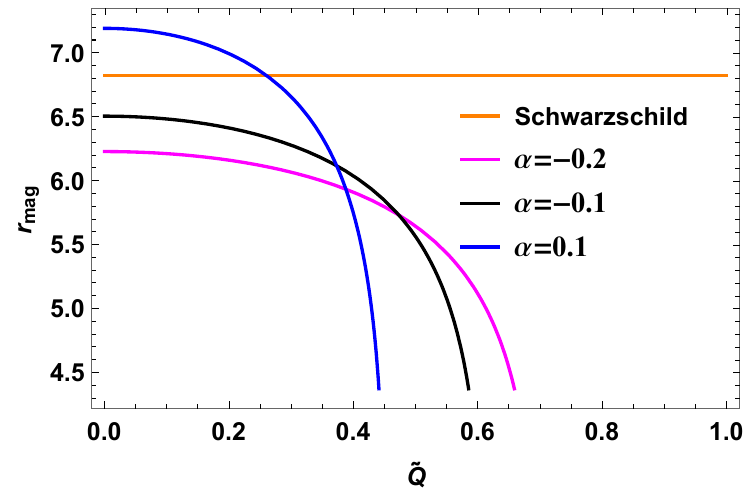}&
\includegraphics[width=0.4\columnwidth]{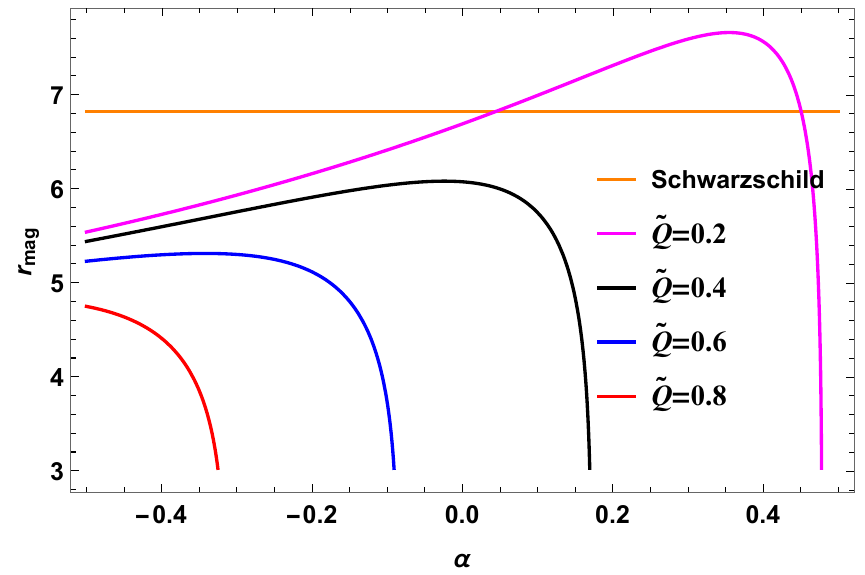}
\end{tabular}
\caption{Variation of relative magnification with charge $\tq$ (left panel) and LV parameter $\a$ (right panel).}\label{rmag}
\end{center}
\end{figure}
Unlike previous cases, here the relative magnification is always smaller than an RN or \s BH when the LV parameter is negative. For its positive values, there exist values of charge that equate $r_{\text{mag}}$ for a QKR BH with that for a \s BH. Few examples of such $(\a,~\tq)$ combinations are $(0.450328,~0.2)$ and $(0.1,~0.260344)$. Fig. (\r{s}) showcases how variation of LV and charge parameters impacts the angular separation. For both the parameters, $s$ initially increases with increasing parameter value, reaching a maximum, and then sharply decreases.
\begin{figure}[H]
\begin{center}
\begin{tabular}{cc}
\includegraphics[width=0.4\columnwidth]{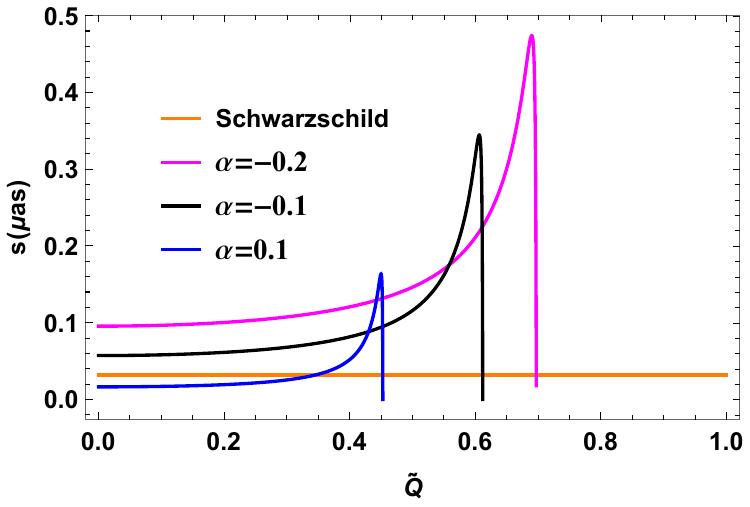}&
\includegraphics[width=0.4\columnwidth]{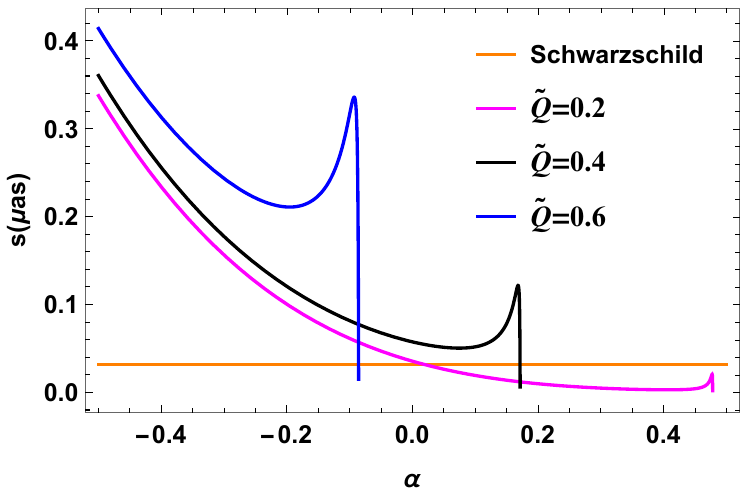}
\end{tabular}
\caption{Variation of angular separation with charge $\tq$ (left panel) and LV parameter $\a$ (right panel). Here, we have modelled SMBH $SgrA^*$ as a QKR BH.}\label{s}
\end{center}
\end{figure}
Fig. (\r{tin}) evidently showcases the adverse impact of both parameters $\a$ and $\tq$ on the angular size $\theta_\infty$. 
\begin{figure}[H]
\begin{center}
\begin{tabular}{cc}
\includegraphics[width=0.4\columnwidth]{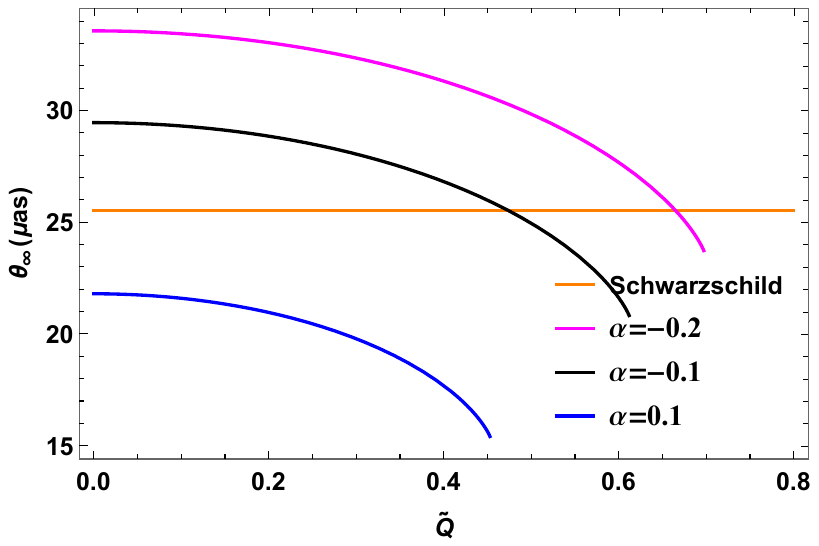}&
\includegraphics[width=0.4\columnwidth]{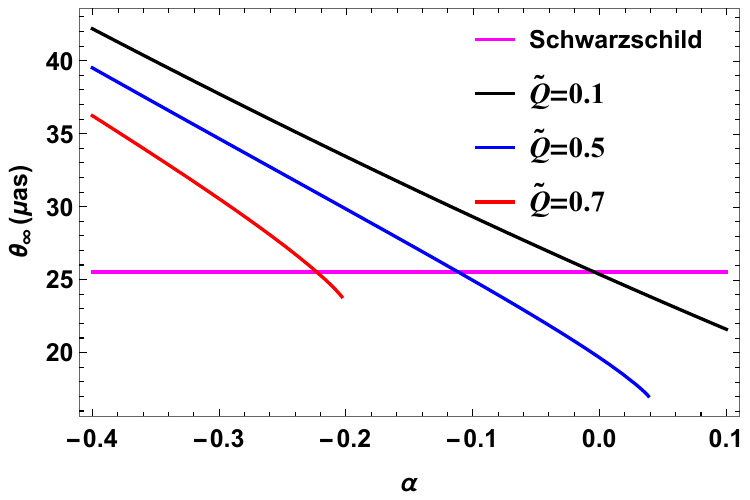}
\end{tabular}
\caption{Variation of angular radius $\theta_\infty$ with charge $\tq$ (left panel) and LV parameter $\a$ (right panel). Here, we have modelled SMBH $SgrA^*$ as a QKR BH.}\label{tin}
\end{center}
\end{figure}
It also reveals that a QKR BH with a positive LV parameter has a smaller shadow than a \s BH. Similar to the previously mentioned cases, for combinations such as $(-0.1,~0.473451)$ and $(-0.2,~0.664742)$, QKR and \s BHs cast shadows of equal size. To accentuate our qualitative evaluation, we tabulate numerical values of lensing observables in (\r{val}).
\begin{table}[H]
\begin{centering}
\setlength{\tabcolsep}{0pt}
\begin{tabular*}{\textwidth}{@{\extracolsep{\fill}\quad}lccccccc }
\hline\hline
\multicolumn{2}{c}{}&
\multicolumn{2}{c}{Sgr A*}&
\multicolumn{2}{c}{M87*}& \\
{$\alpha$ }& {$\tq$} & {$\theta_\infty $ ($\mu$as)} & {$s$ ($\mu$as) } & {$\theta_\infty $ ($\mu$as)} & {$s$ ($\mu$as) } & {$r_{mag} $ }
\\ \hline
\hline
\multirow{5}{*}{-0.1}&$0$ & $29.4534$ & $0.0571544$ & $22.82$ & $22.82$ & $6.50441$ \\
&$0.1$ & $29.305$ & $0.0581219$ & $22.705$ & $22.705$ & $6.48236$ \\
&$0.2$ & $28.8488$ & $0.0613273$ & $22.3516$ & $22.3516$ & $6.4119$ \\
&$0.3$ & $28.0471$ & $0.0679566$ & $21.7304$ & $21.7304$ & $6.27721$ \\
&$0.4$ & $26.8155$ & $0.08153$ & $20.7762$ & $20.7762$ & $6.03719$ \\
\hline
\multirow{5}{*}{0}&$0$ & $25.5298$ & $0.0319504$ & $19.78$ & $19.78$ & $6.82188$ \\
&$0.1$ & $25.3582$ & $0.0327412$ & $19.6471$ & $19.6471$ & $6.79094$ \\
&$0.2$ & $24.8263$ & $0.0354663$ & $19.235$ & $19.235$ & $6.68985$ \\
&$0.3$ & $23.8718$ & $0.0416744$ & $18.4954$ & $18.4954$ & $6.48575$ \\
&$0.4$ & $22.3354$ & $0.0574832$ & $17.305$ & $17.305$ & $6.07378$ \\
\hline
\multirow{5}{*}{0.1}&$0$ & $21.7977$ & $0.0163845$ & $16.8885$ & $16.8885$ & $7.19089$ \\
&$0.1$ & $21.5962$ & $0.017002$ & $16.7323$ & $16.7323$ & $7.14581$ \\
&$0.2$ & $20.963$ & $0.0192708$ & $16.2417$ & $16.2417$ & $6.99329$ \\
&$0.3$ & $19.7854$ & $0.0254174$ & $15.3293$ & $15.3293$ & $6.65501$ \\
&$0.4$ & $17.686$ & $0.0518673$ & $13.7028$ & $13.7028$ & $5.74464$ \\
\hline\hline
\end{tabular*}
\end{centering}
\caption{Strong-lensing observables for supermassive black holes $Sgr A*$ and $M87^*$. \label{val}}
\end{table}
Table (\r{val}) reinforces our findings from the qualitative study. With the pre-requisite platform being set, we now move on to constrain parameters $\a$ and $Q$ utilizing experimental observations related to SMBHs $Sgr A*$ and $M87^*$ (\r{data}) by modeling them as QKR BH.\\
\textbf{Constrain from $\text{SgrA}^*$}: Within $1\sigma$ bounds, the angular diameter for the SMBH $SgrA^*$ lies within the range $ (41.7,~55.7)~\mu as$. Fig. (\r{sgra}) displays variation of $\theta_d$ with $\a$ and $Q$. For our model to be consistent with this observation, we obtain $\a \in [-0.0597061,~0.126279]$ with no bound on charge. However, the condition (\r{qb}) yields $Q \le 0.545441M$ for existence of BH.
\begin{figure}[H]
\begin{center}
\begin{tabular}{cccc}
\includegraphics[width=0.4\columnwidth]{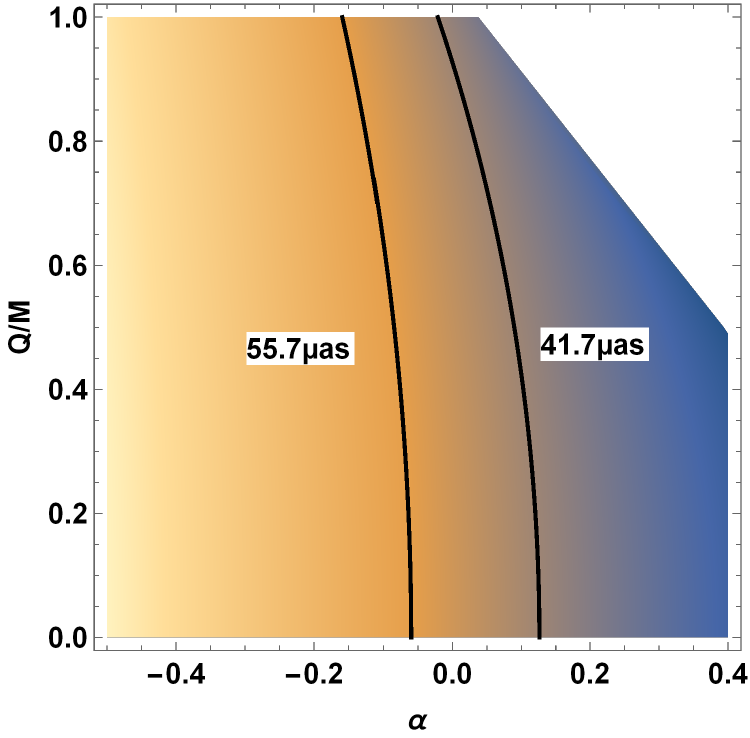}&
\includegraphics[width=0.045\columnwidth]{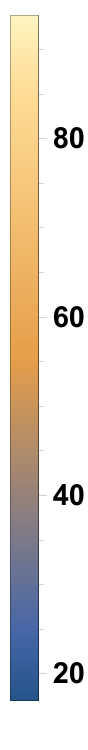}
\end{tabular}
\caption{Variation of angular diameter $\theta_d$ for the SMBH $SgrA^*$ with LV parameter $\a$ and charge $Q$. The upper solid black line corresponds to the lower $1\sigma$ bound, and the lower one is for the upper $1 \sigma$ bound. }\label{sgra}
\end{center}
\end{figure}
\textbf{Constrain from $\text{M87}^*$}: Within $1\sigma$ bounds, the angular diameter for the SMBH $M87^*$ lies within the range $ (39,~45)~\mu as$. Fig. (\r{m87}) displays variation of $\theta_d$ with $\a$ and $Q$. For our model to commensurate with this observation, we obtain $\a \in [-0.0896928,~0.00945968]$ with no bound on charge. However, if we consider the condition (\r{qb}), then we must have $Q \le 0.568756M$.
\begin{figure}[H]
\begin{center}
\begin{tabular}{cccc}
\includegraphics[width=0.4\columnwidth]{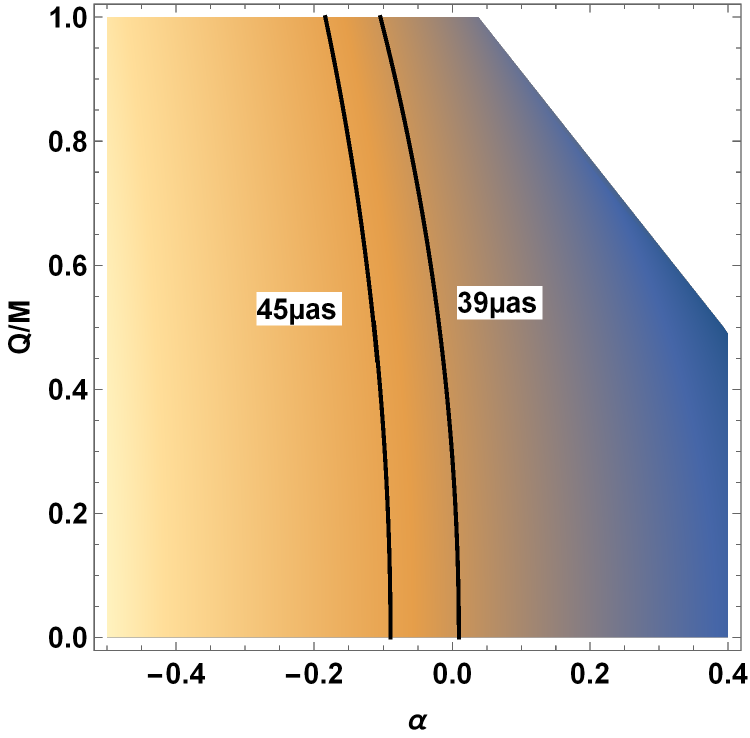}&
\includegraphics[width=0.045\columnwidth]{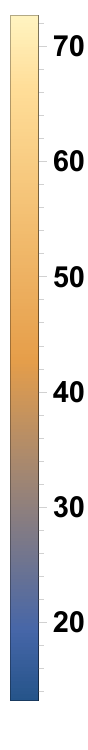}
\end{tabular}
\caption{Variation of angular diameter $\theta_d$ for the SMBH $M87^*$ with LV parameter $\a$ and charge $Q$. The upper solid black line corresponds to the lower $1\sigma$ bound, and the lower one is for the upper $1 \sigma$ bound. }\label{m87}
\end{center}
\end{figure}
Bounds thus gleaned clearly exhibit the viability of our model as an SMBH. The next section will put our model to another test for gauging its commensurability with observed results.
\section{Testing qkr bh with qpo observations}
QPOs are commonly observed as narrow peaks in the X-ray power density spectra of binaries consisting of neutron stars or BHs. These oscillations are related to the motion of test particles near innermost circular orbit (ISCO) and are therefore functions of the curvature of spacetime alone. To this end, we write the Lagrangian for the motion of a test particle on the equatorial plane as
\be
\mathscr{L}=\f{1}{2}\lt(g_{tt} \dot{t}^2+g_{rr}\dot{r}^2+g_{\phi\phi} \dot{\phi}^2\rt),
\ee
where $\gtt=-\fr$, $\grr=1/\fr$, and $\gth=\hr=r^2$. Similar to the photon motion, $\e$ and $\l$ are conserved here given by
\ba\nn
\e&=&-p_{t}=-\f{\pt \mathscr{L}}{\pt \dot{t}}=-g_{tt} \dot{t}=\fr \dot{t},\\\nn
\l&=&p_{\phi}=\f{\pt \mathscr{L}}{\pt \dot{\phi}}=g_{\phi\phi} \dot{\phi}=\hr \dot{\phi},
\label{pt}
\ea
with one dissimilarity: here $\e$ and $\l$ represent, respectively, specific energy and momentum of the test particle. This, conjoined with the condition $u_{\mu}u^{\mu}=-1$ followed by the four-velocity of test particles, allows us to obtain the differential equation of motion as
\ba\nn
\grr \dot{r}^2&=&-\f{1}{\gtt}\lt[\e^2+\gtt(1+\f{\l^2}{\gth})\rt]\\\nn
&=&-\f{1}{\gtt}\lt[\e^2-\mc{V}_{\text{eff}}\rt].
\ea
Here, the potential $\mc{V}_{\text{eff}}=-\gtt(1+\f{\l^2}{\gth})$ governs the motion of test particles. For circular motion, we have
\be
\veff(r_0)=\e^2, \quad \quad \f{\pt \veff}{\pt r}|_{r=r_0}=0,
\ee
leading to following expressions of $\e$ and $\l$ as
\be
\l^2=\f{-\gtt'\gth^2}{\gth\gtt'-\gtt\gth'},\quad \quad \quad \quad \e^2=\f{\gtt^2 \gth'}{\gth\gtt'-\gtt\gth'},
\label{conserved}
\ee
where $'$ represents partial differentiation with respect to $r$. We obtain the position of ISCO ($r_{isco}$) by imposing the condition $\f{\pt^2 \veff}{\pt r^2}|_{r=r_{isco}}=0$ that leads to the following equation of $r$ whose solution provides $r_{isco}$:
\be
-9 (\alpha -1)^2 M Q^2 r+(\alpha -1)^3 M r^2 (6 (\alpha -1) M+r)+4 Q^4=0.
\ee
Fig. (\r{risco}) showcases a qualitative overview of the impact parameters $\a$ and $Q$ have on the ISCO radius. As on previous occasions, $r_{isco}$ shows the adverse impact of increasing either parameter. For a negative value of the LV parameter, the radius of the last stable orbit is larger for a QKR BH than for an RN BH. The situation is reversed when it comes to the positive value of $\a$.
\begin{figure}[H]
\begin{center}
\begin{tabular}{cc}
\includegraphics[width=0.4\columnwidth]{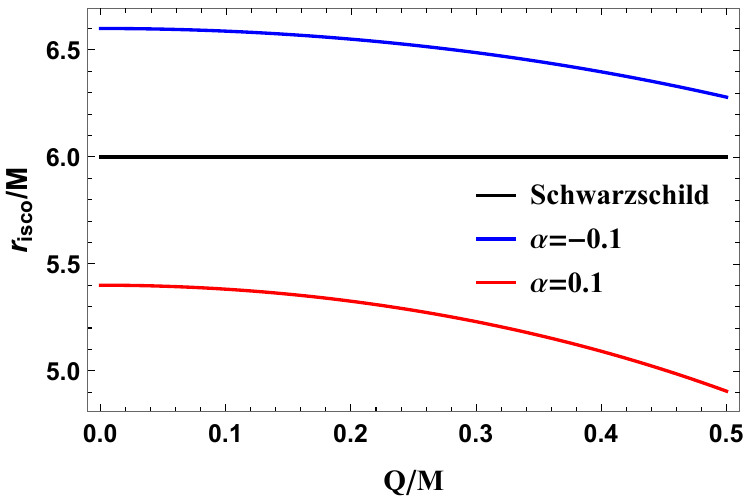}&
\includegraphics[width=0.4\columnwidth]{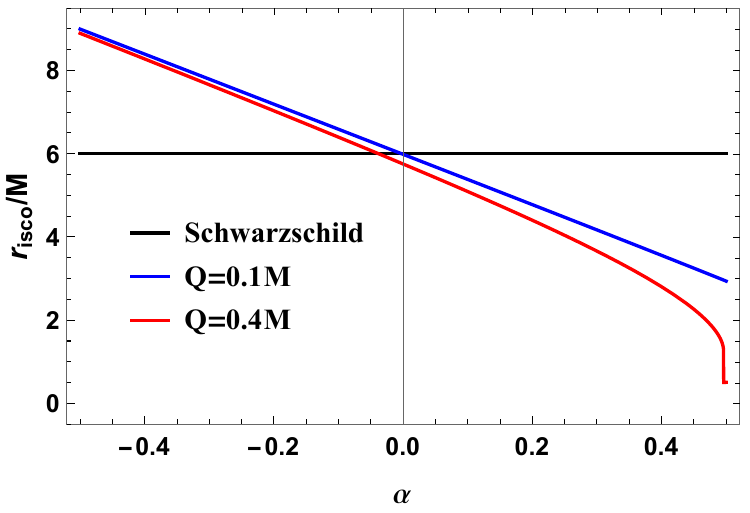}
\end{tabular}
\caption{Variation of ISCO radius with charge $Q$ (left panel) and with LV parameter $\a$ (right panel). }\label{risco}
\end{center}
\end{figure}
We now move on to obtain epicyclic frequencies in the background of a QKR BH. Epicyclic oscillations occur when a test particle is perturbed from its stable circular orbit. In such a scenario, the particle undergoes two motions in perpendicular directions: one radial in the equatorial plane and another latitudinal normal to the plane. If the radial and latitudinal perturbations are, respectively, $\delta r$ and $\delta \theta$, then for small perturbations, the two mutually perpendicular motions satisfy the following differential equations:
\ba
\delta \ddot{r}+\Omega_{r}^2 \delta r=0 \text{and} \quad \quad \delta \ddot{\theta}+\Omega_{\theta}^2 \delta \theta=0,
\ea
where $\Omega_{r}$ and $\Omega_{\theta}$ are, respectively, radial and angular frequencies defined locally, and the dot above connotes differentiation against proper time. To extract expressions for epicyclic frequencies, we separate the Hamiltonian into dynamical part $\mc{H}_{dyn}$ and potential part $\mc{H}_{pot}$ as
\ba\nn
\mc{H}_{dyn}&=&\f{1}{2}\lt(\f{p_{r}^2}{\grr}+\f{p_{\theta}^2}{g_{\theta \theta}}\rt),\\
\mc{H}_{pot}&=&\f{1}{2}\lt(\f{\e^2}{\gtt}+\f{\l^2}{\gth}+1\rt).
\label{hpot}
\ea
The local frequencies are governed by the potential part $\mc{H}_{dyn}$ of the Hamiltonian. Following expressions provide the epicyclic frequencies $\Omega_{r}$ and $\Omega_{\theta}$ as \c{sanjar, vrba}
\ba\nn
\Omega_{r}^2&=&\f{1}{\grr}\f{\pt^2 \mc{H}_{pot}}{\pt r^2},\\\nn
\Omega_{\theta}^2&=&\f{1}{\grr}\f{\pt^2 \mc{H}_{pot}}{\pt \theta^2}.
\label{Omega}
\ea
Since QPOs are measured far away from the celestial body, we need to convert the above locally measured frequencies $\Omega$ into frequencies $\omega$ measured at spatial infinity. The required relation is \c{sanjar, vrba}
\be
\omega \rightarrow \f{\Omega}{-g^{tt}\e}.
\label{omega}
\ee
Eq. (\r{Omega}) and (\r{omega}) together with Eq. (\r{conserved}) provide the desired expressions for radial and latitudinal frequencies as:
\ba
\nu_r&=&\f{1}{2\pi} \sqrt{\f{2\gth \gtt'^2-2\gtt \gtt' \gth-\gtt \gth \gtt''}{2\gtt \grr \gth}+\f{\gtt' \gth''}{2\grr \gth'}},\\
\nu_\theta&=&\nu_\phi=\f{1}{2\pi}\sqrt{-\f{\gtt'}{\gth'}}.
\label{nu}
\ea
In terms of model parameters and mass of the BH, the above expressions become
\ba
\nu_r&=&\f{1}{2\pi} \sqrt{\frac{M \left(Q^2-(\alpha -1) r (2 (\alpha -1) M+r)\right)}{(\alpha -1)^2 r^5}},\\
\nu_\theta&=&\f{1}{2\pi}\sqrt{\frac{(\alpha -1)^2 M r-Q^2}{(\alpha -1)^2 r^4}}.
\ea
To have frequencies in Hz, we use the transformation $\nu \rightarrow \nu \f{c^3}{GM}$, where $c$ and $G$ are, respectively, the velocity of light in free space and the gravitational constant. With the theoretical background laid, we now test our model against the observational frequencies for the microquasars $GRO J1655-40$ and $XTE J1550-564$, whose power spectra show two sharp peaks in the $3:2$ ratio. Their lower $\nu_L$ and upper $\nu_U$ frequencies together with their masses are given as \c{qpo40, qpo401, qpo564}:
\ba\nn
\tx{GRO J1655-40:}~~ \f{M}{M_{\odot}}=6.30\pm 0.27,\quad \nu_{U}=450\pm 3 Hz, \quad \nu_{L}=300\pm 5 Hz,\\
\tx{XTE J1550-564:} ~~\f{M}{M_{\odot}}=9.10\pm 0.60,\quad \nu_{U}=276\pm 3 Hz, \quad \nu_{L}=184\pm 5 Hz.\label{data1}
\ea
Among various theorietical models that try to explain observed twin peaks in the power spectrum, one plausible explanation is resonance between radial and latitudinal oscillations that occur near ISCO. In resonance model, nonlinear coupling between the two mutually perpendicular oscillations is considered to be responsible for the observed QPOs \c{resonance, resonance1}. The observed ratio between $\nu_U$ and $\nu_L$ also points towards a possible resonance. We follow the forced resonance model where the upper and lower QPO frequencies are given by \c{forced}:
\be
\nu_{L}=\nu_\theta \quad \quad \tx{and} \quad \quad \nu_{U}=\nu_{r}+\nu_{\theta}.
\ee
In Fig. (\r{nu}), the upper frequency analyzed near the ISCO radius is plotted as a function of the BH mass $ M$ for various values of the LV and charge parameters. Fig. (\r{nl}) illustrates the same for the lower frequency. As a BH with a larger mass has a stronger gravitational pull, the ISCO radius increases with $M$. Since it is near ISCO where QPO occurs, QPO frequencies decrease with $M$, which is evident from Fig. (\r{nu}) and (\r{nl}). These figures also exhibit the adverse impact of increasing $\a$, but the favorable effect of increasing $Q$ on the QPO frequency.
\begin{figure}[H]
\begin{center}
\begin{tabular}{cc}
\includegraphics[width=0.4\columnwidth]{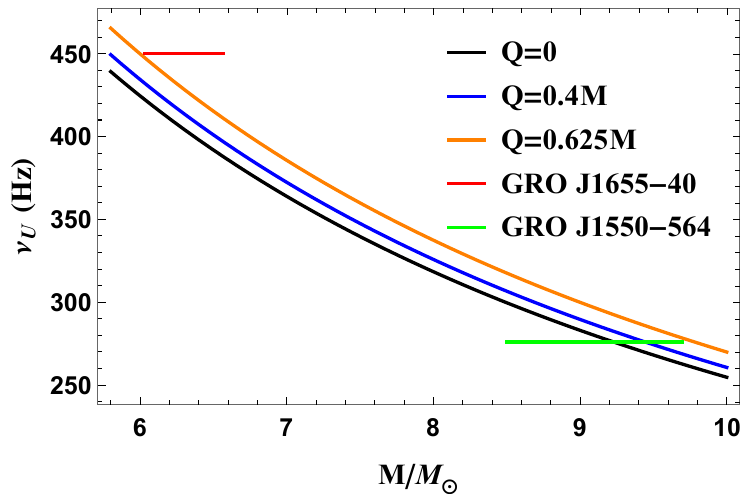}&
\includegraphics[width=0.4\columnwidth]{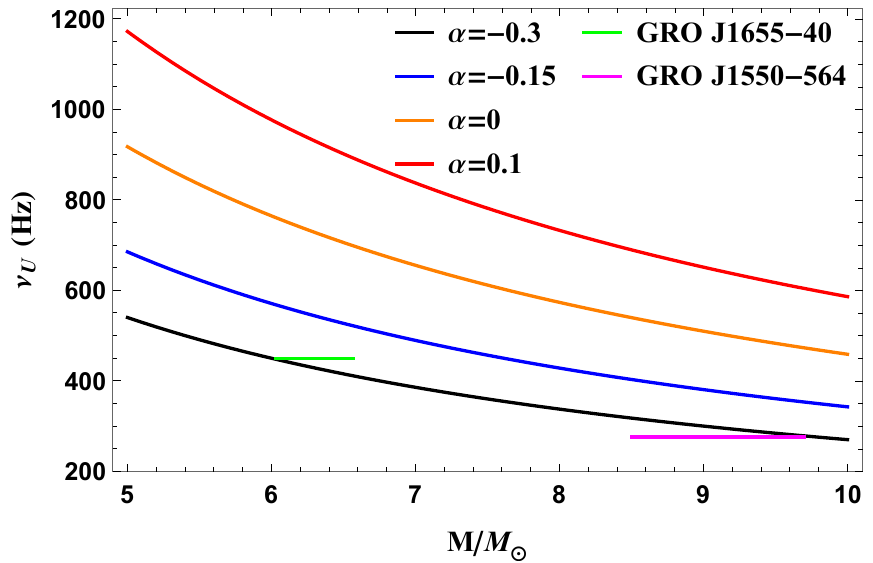}
\end{tabular}
\caption{Variation of upper frequency $\nu_{U}$ with charge $Q$ (left panel) at a fixed value of $\a=-0.3$ and with LV parameter $\a$ (right panel) at a fixed value of $Q=0.625M$. }\label{nu}
\end{center}
\end{figure}
\begin{figure}[H]
\begin{center}
\begin{tabular}{cc}
\includegraphics[width=0.4\columnwidth]{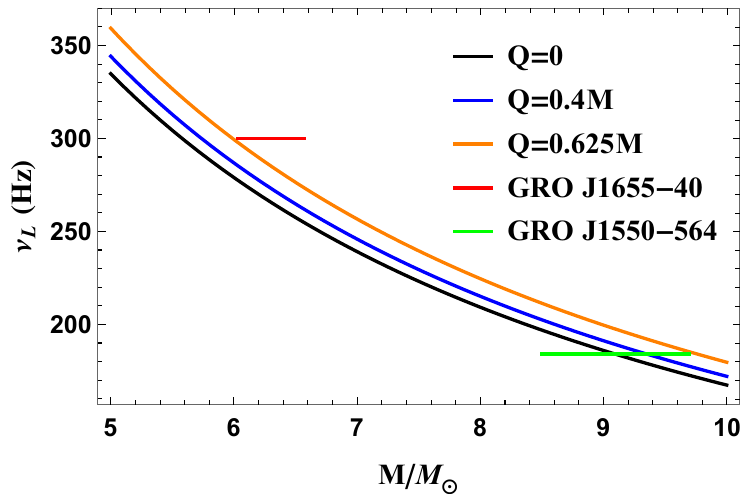}&
\includegraphics[width=0.4\columnwidth]{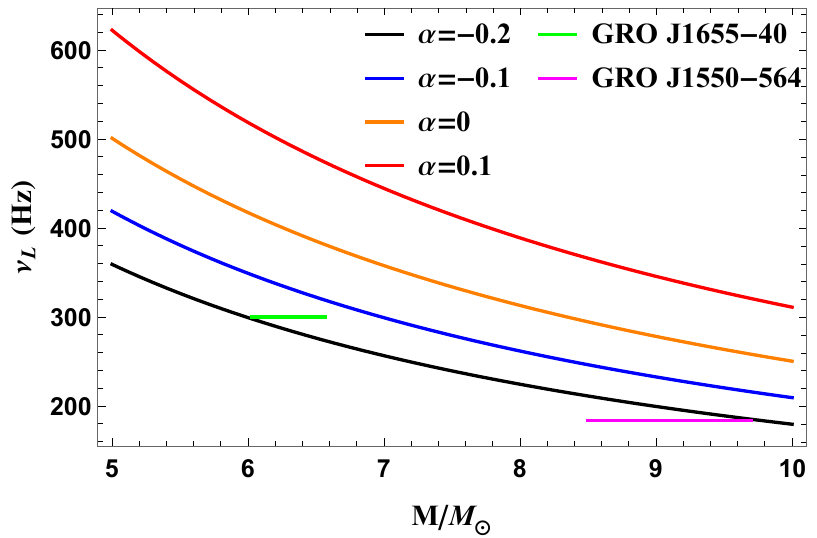}
\end{tabular}
\caption{Variation of lower frequency $\nu_{L}$ with charge $Q$ (left panel) at a fixed value of $\a=-0.2$ and with LV parameter $\a$ (right panel) at a fixed value of $Q=0.625M$. }\label{nl}
\end{center}
\end{figure}
Fig. (\r{final}) showcases the variation of lower and upper frequencies near $r_{isco}$ as a function of $M/M_{\odot}$ for the best-fit values of LV and charge parameters.
\begin{figure}[H]
\begin{center}
\begin{tabular}{c}
\includegraphics[width=0.5\columnwidth]{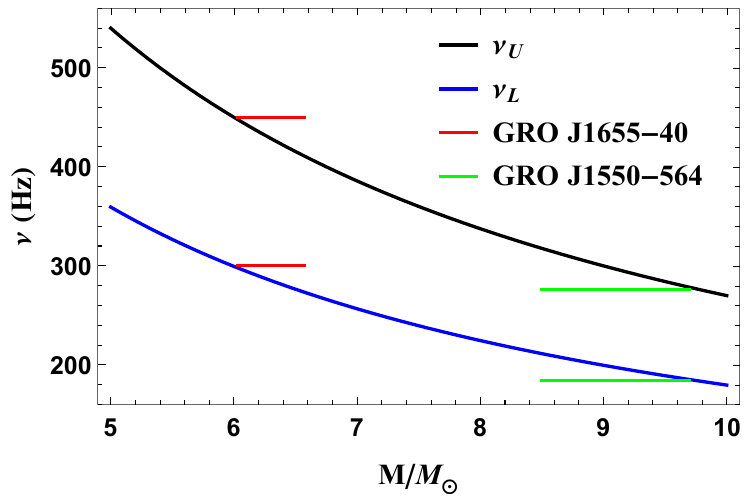}
\end{tabular}
\caption{Variation of lower and upper frequencies $\nu$ for best-fit values. }\label{final}
\end{center}
\end{figure}
The best-fit values of LV and charge parameters for the upper frequency are $\a=-0.3^{+0.012}_{-0.012}$ and $Q/M=0.625^{+0.086}_{-0.096}$. These values for the lower frequency are $\a=-0.2^{+0.011}_{-0.011}$ and $Q/M=0.625^{+0.060}_{-0.073}$. While results pertaining to the shadow of SMBHs $M87^*$ and $SgrA^*$ could not bound the electric charge, observations related to HF QPOs were able to constrain it. These results once again showcase the feasibility of our model.
\section{conclusion}
 This manuscript is primarily focused on constraining LV and charge parameters for QKR BHs, where we have utilized experimental observations of two phenomena: one related to SGL observables of SMBHs $M87^*$ and $SgrA^*$, and the other to QPOs of microquasars. While the former concerns the photon trajectory near the BH, the latter arises from the motion of test particles near the ISCO. Since both these motions occur very close to BH, they carry signatures of the intrinsic nature of the underlying spacetime. Our study revealed cases where competing effects of LV and charge cancel each other, thereby leaving event horizon, shadow radius, and ISCO radius unchanged from that of a \s BH. In all such cases, the LV parameter is always negative.\\
Our inquisition into SGL sheds light on the interplay between charge and LV. While a QKR BH with a negative LV parameter value pulls photons more strongly than an RN or \s BH, thereby resulting in greater deviation from their original path, the situation is reversed for positive values of $\a$. This repeats itself when we consider the angular diameter of BH shadows. Having explored the variation of lensing observables against charge and LV parameters, we tested our BH metric against observations related to SMBHs $M87^*$ and $SgrA^*$ to glean bounds on $\a$ and $Q$. Bounds obtained from $M87^*$ shadow observation is $\a \in [-0.0896928,~0.00945968]$ whereas observations pertaining to $SgrA^*$ yielded $\a \in [-0.0597061,~0.126279]$ within $1\sigma$. No bound for charge could be extracted from these observations. However, by employing the condition necessary for the existence of BH, we have constrained the charge as well. \\
Observations of twin peaks in the X-ray power density spectra of binaries, such as BHs and neutron stars, provide an alternative and complementary avenue for examining the commensurability of a proposed model. We have considered observed QPOs of microquasars $GRO J1655-40$ and $XTE J1550-564$. Among the various models that attempt to explain observed QPOs, we chose the forced resonance model. An increasing mass of the BH is found to impact QPO frequencies adversely. This occurs because increasing mass enhances the gravitational pull of the BH, thereby pushing the ISCO, where the QPO happens, further from the BH, leading to a decrease in its frequency. We obtained best-fit values of $\a$ and $Q$ for both lower and upper frequencies. The best-fit values of LV and charge parameters for the upper frequency are $\a=-0.3^{+0.012}_{-0.012}$ and $Q/M=0.625^{+0.086}_{-0.096}$. These values for the lower frequency are $\a=-0.2^{+0.011}_{-0.011}$ and $Q/M=0.625^{+0.060}_{-0.073}$.\\
Our results shed new light on the intricate interplay between charge and LV. It enhances our understanding of how the combined effects of charge and LV influence the motion of photons and matter. The results reported here evidently enable us to differentiate a QKR BH from an \s or RN BH. Observations with higher precision from next-generation telescopes are expected to place tighter constraints on model parameters and provide deeper insight into strong-gravity phenomena.


\begin{thebibliography}{the}
\bibitem{Kostelecky1989a}
V.A.~Kostelecky and S.~Samuel, Spontaneous breaking of Lorentz symmetry in string theory, Phys. Rev. D 39 (1989) 683.

\bibitem{Alfaro2002}
J.~Alfaro et al., Loop quantum gravity and light propagation, Phys. Rev. D 65 (2002) 103509.

\bibitem{Horava2009a}
P.~Horava, Quantum Gravity at a Lifshitz Point, Phys. Rev. D 79 (2009) 084008.

\bibitem{Carroll2001}
S.M.~Carroll et al., Noncommutative field theory and Lorentz violation, Phys. Rev. Lett. 87 (2001) 141601.

\bibitem{Jacobson2001}
T.~Jacobson and D.~Mattingly, Gravity with a dynamical preferred frame, Phys. Rev. D 64 (2001) 024028.

\bibitem{Dubovsky2005}
S.L.~Dubovsky et al., Massive graviton as a testable cold dark matter candidate, Phys. Rev. Lett. 94 (2005) 181102.

\bibitem{Bengochea2009}
G.R.~Bengochea and R.~Ferraro, Dark torsion as the cosmic speed-up, 
Phys. Rev. 79 (2009) 124019.

\bibitem{Cohen2006}
A.G.~Cohen and S.L.~Glashow, Very special relativity, Phys. Rev. Lett. 97 (2006) 021601.

\bibitem{Kostelecky2004a}
V.A.~Kostelecky, Gravity, Lorentz violation, and the standard model, Phys. Rev. D 69 (2004) 105009.

\bibitem{Kostelecky1989}
V.A.~Kostelecky and S.~Samuel, Gravitational Phenomenology in Higher Dimensional Theories and Strings, 
Phys. Rev. D 40 (1989) 1886.

\bibitem{Kostelecky1989b}
V.A.~Kostelecky and S.~Samuel, Phenomenological gravitational constraints on strings and higher dimensional theories, 
Phys. Rev. Lett. 63 (1989) 224.

\bibitem{Bailey2006}
Q.G.~Bailey and V.A.~Kostelecky, Signals for Lorentz violation in post-Newtonian gravity, 
Phys. Rev. D 74 (2006) 045001.

\bibitem{Bluhm2008a}
R.~Bluhm et al., Constraints and stability in vector theories with spontaneous Lorentz violation, Phys. Rev. D 77 (2008) 125007.

\bt{action1} B. Altschul et al., Lorentz violation with
an antisymmetric tensor, Phys. Rev. D 81, 065028 (2010). 

\bibitem{bm} R. Casana et al., An exact Schwarzschild-like solution in a bumblebee gravity model, Phys. Rev. D 97, 104001 (2018).

\bt{action2} L.A. Lessa et al., Modified
black hole solution with a background Kalb-Ramond field, Eur. Phys. J. C 80 (2020) 335.

\bibitem{kr} Ke Yang et al., Static and spherically symmetric black holes in gravity with a background Kalb-Ramond field: Phys. Rev. D 108, 124004 (2023).

\bibitem{Ovgun2018}
A.~Ovg{\"u}n et al., Gravitational lensing under the effect of Weyl and bumblebee gravities: Applications of Gauss-Bonnet theorem, Annals Phys. 399 (2018) 193.

\bibitem{Kanzi2019}
S.~Kanzi and {\.I}.~Sakall{\i}, GUP modified hawking radiation in bumblebee gravity, 
Nucl. Phys. B 946 (2019) 114703.

\bibitem{Yang2019b}
R.-J.~Yang et al., Effects of Lorentz breaking on the accretion onto a Schwarzschild-like black hole, 
Commun. Theor. Phys. 71 (2019) 568.

\bibitem{Cai2022a}
Z.~Cai and R.-J.~Yang, Accretion of the Vlasov gas onto a Schwarzschild-like black hole,  
Phys. Dark Univ. 42 (2023) 101292.

\bibitem{Oliveira2021}
R.~Oliveira et al., Quasinormal frequencies for a black hole in a bumblebee gravity, 
EPL 135 (2021) 10003.

\bibitem{Maluf2021}
R.V.~Maluf and J.C.S.~Neves, Black holes with a cosmological constant in bumblebee gravity, 
Phys. Rev. D 103 (2021) 044002.

\bibitem{Xu2023}
R.~Xu et al., Static spherical vacuum solutions in the bumblebee gravity model, 
Phys. Rev. D 107 (2023) 024011.

\bibitem{Mai2023}
Z.-F.~Mai et al., Extended thermodynamics of the bumblebee black holes, 
Phys. Rev. D 108 (2023) 024004.

\bibitem{Xu2023a}
R.~Xu et al., Bumblebee black holes in light of event horizon telescope observations, 
Astrophys. J. 945 (2023) 148.

\bibitem{Liang2023}
D.~Liang et al., Probing vector hair of black holes with extreme-mass-ratio inspirals, 
Phys. Rev. D 107 (2023) 044053.

\bibitem{Ding2020a}
C.~Ding et al., Exact Kerr-like solution and its shadow in a gravity model with spontaneous Lorentz symmetry breaking, Eur. Phys. J. C 80 (2020) 178.

\bibitem{Ding2021a}
C.~Ding and X.~Chen, Slowly rotating Einstein-bumblebee black hole solution and its greybody factor in a Lorentz violation model, 
Chin. Phys. C 45 (2021) 025106.

\bibitem{Wang2022}
H.-M.~Wang and S.-W.~Wei, Shadow cast by Kerr-like black hole in the presence of plasma in Einstein-bumblebee gravity, 
Eur. Phys. J. Plus 137 (2022) 571.

\bibitem{Liu2019}
C.~Liu et al., Thin accretion disk around a rotating Kerr-like black hole in Einstein-bumblebee gravity model, 
[arXiv:1910.13259].

\bibitem{Liu2023}
W.~Liu et al., QNMs of slowly rotating Einstein-Bumblebee black hole, 
Eur. Phys. J. C 83 (2023) 83.

\bibitem{Wang2022a}
Z.~Wang et al., Constraint on parameters of a rotating black hole in Einstein-bumblebee theory by quasi-periodic oscillations, Eur. Phys. J. C 82 (2022) 528.

\bibitem{Ding2023}
C.~Ding et al., Rotating BTZ-like black hole and central charges in Einstein-bumblebee gravity, Eur. Phys. J. C 83 (2023) 573.

\bibitem{Chen2023}
C.~Chen et al., Quasinormal modes of a scalar perturbation around a rotating BTZ-like black hole in Einstein-bumblebee gravity, Phys. Lett. B 846 (2023) 138186.

\bibitem{Gullu2022}
{\.I}.~G{\"u}ll{\"u} and A.~{\"O}vg{\"u}n, Schwarzschild-like black hole with a topological defect in bumblebee gravity, Annals Phys. 436 (2022) 168721.

\bibitem{Zhang2023}
X.~Zhang et al., Quasinormal modes and late time tails of perturbation fields on a Schwarzschild-like black hole with a global monopole in the Einstein-bumblebee theory, Sci. China Phys. Mech. Astron. 66 (2023) 100411.

\bibitem{Lin2023}
R.-H.~Lin et al., Quasinormal modes of the spherical bumblebee black holes with a global monopole, Eur. Phys. J. C 83 (2023) 720.

\bibitem{Ding2021}
C.~Ding et al., Einstein-Gauss-Bonnet gravity coupled to bumblebee field in four dimensional spacetime, Nucl. Phys. B 975 (2022) 115688.

\bibitem{Jha2021}
S.K.~Jha and A.~Rahaman, Bumblebee gravity with a Kerr-Sen-like solution and its Shadow, Eur. Phys. J. C 81 (2021) 345.

\bibitem{Ding2023a}
C.~Ding et al., High dimensional AdS-like black hole and phase transition in Einstein-bumblebee gravity, Chin. Phys. C 47 (2023) 045102.

\bibitem{Ovgun2019}
A.~{\"O}vg{\"u}n et al., Exact traversable wormhole solution in bumblebee gravity, 
Phys. Rev. D 99 (2019) 024042.

\bibitem{Liang2022}
D.~Liang et al., Polarizations of Gravitational Waves in the Bumblebee Gravity Model,
Phys. Rev. D 106 (2022) 124019. 

\bibitem{Amarilo2023}
K.M.~Amarilo et al., Gravitational waves effects in a Lorentz-violating scenario,
[arXiv:2307.10937].


\bibitem{Kalb1974}
M.~Kalb and P.~Ramond, Classical direct interstring action, 
Phys. Rev. D 9 (1974) 2273.

\bibitem{Kao1996}
W.F.~Kao et al., Induced Einstein-Kalb-Ramond theory and the black hole, 
Phys. Rev. D 53 (1996) 2244.

\bibitem{Kar2003}
S.~Kar et al., Static spherisymmetric solutions, gravitational lensing and perihelion precession in Einstein-Kalb-Ramond theory, 
Phys. Rev. D 67 (2003) 044005.

\bibitem{Chakraborty2017}
S.~Chakraborty and S.~SenGupta, Strong gravitational lensing \textemdash{} a probe for extra dimensions and Kalb-Ramond field, 
JCAP 07 (2017) 045.

\bibitem{Nair2022}
K.K.~Nair and A.M.~Thomas, Kalb-Ramond field-induced cosmological bounce in generalized teleparallel gravity, 
Phys. Rev. D 105 (2022) 103505.

\bibitem{Fu2012}
C.-E.~Fu et al., Q-form fields on p-branes, 
JHEP 10 (2012) 060.

\bibitem{Chakraborty2016}
S.~Chakraborty and S.~SenGupta, Solutions on a brane in a bulk spacetime with Kalb-Ramond field, 
Annals Phys. 367 (2016) 258.


\bibitem{Atamurotov2022}
F.~Atamurotov et al., Particle dynamics and gravitational weak lensing around black hole in the Kalb-Ramond gravity, 
Eur. Phys. J. C 82 (2022) 659.

\bibitem{Kumar2020c}
R.~Kumar et al., Gravitational deflection of light and shadow cast by rotating Kalb-Ramond black holes, 
Phys. Rev. D 101 (2020) 104001.

\bibitem{Lessa2021}
L.A.~Lessa et al., Traversable wormhole solution with a background Kalb-Ramond field, 
Annals Phys. 433 (2021) 168604.

\bt{jha25}
S.Kr.~Jha, Black hole surrounded by perfect fluid dark matter with
a background Kalb-Ramond field, JCAP 09(2025)069.

\bibitem{Maluf2022}
R.V.~Maluf and C.R.~Muniz, Exact solution for a traversable wormhole in a curvature-coupled antisymmetric background field, 
Eur. Phys. J. C 82 (2022) 445.
\bibitem{Maluf2022a}
R.V.~Maluf and J.C.S.~Neves, Bianchi type I cosmology with a Kalb-Ramond background field, 
Eur. Phys. J. C 82 (2022) 135.
\bibitem{qkr} Z.~Q.~Duan et al., Electrically charged black holes in gravity with a background Kalb-Ramond field, Eur. Phys. J. C 84 (2024), 798.
\bt{dar} C. Darwin, The gravity field of a particle, Proc. R. Soc. A 249 (1959) 180.
\bibitem{vir} K. S. Virbhadra and G. F. R. Ellis, Schwarzschild black hole lensing, Phys. Rev. D 62 (2000) 084003.
\bibitem{BOZZA}V. Bozza, S. Capozziello, G. Iovane G. Scarpetta, Strong Field Limit of Black Hole Gravitational Lensing, Gen. Rel. Grav. 33 (2001) 1535.
\bibitem{BOZZA1}V. Bozza, Gravitational lensing in the strong field limit, Phys. Rev. D 66 (2002) 103001.
\bibitem{BOZZA2}V. Bozza and L. Mancini, Time Delay in Black Hole Gravitational Lensing as a Distance Estimator, Gen. Rel. Grav. 36 (2004) 435.
\bt{lens1} E. F. Eiroa and D. F. Torres, Strong field limit analysis of gravitational retro-lensing, Phys. Rev. D 69 (2004) 063004.
\bt{lens2} R. Whisker, Strong gravitational lensing by braneworld black holes, Phys. Rev. D 71 (2005) 064004.
\bt{lens3} E. F. Eiroa, Braneworld black hole gravitational lens: Strong field limit analysis, Phys. Rev. D 71 (2005) 083010.
\bt{lens6} A. Bhadra, Gravitational lensing by a charged black hole of string theory, Phys. Rev. D 67 (2003) 103009.
\bt{lens8} R. Shaikh et al., Analytical approach to strong gravitational lensing from ultracompact objects, Phys. Rev. D 99 (2019) 104040.
\bt{lens9} E. F. Eiroa and C. M. Sendra, Gravitational lensing by a regular black hole, Class. Quant. Grav. 28 (2011) 085008.
\bt{lens11} J. Kumar et al., Testing Strong Gravitational Lensing Effects of Supermassive Compact Objects with Regular Spacetimes, Astrophys. J. 938 (2022) 104.
\bt{lens12} S. K. Jha and A. Rahaman, Strong gravitational lensing in hairy Schwarzschild background, Eur. Phys. J. Plus 138 (2023) 86.
\bt{lens13} F. Feleppa et al., Strong deflection limit analysis of black hole lensing in inhomogeneous plasma, Phys. Rev. D 110 (2024) 064031.
\bt{id1} A. Vachher et al., Probing dark matter via strong gravitational lensing by black holes, Phys. Dark Univ. 44 (2024) 101493.
\bt{id2} N. U.Molla et al., Strong gravitational lensing by $SgrA^*$ and $M87^*$ black holes embedded in dark matter halo exhibiting string cloud and quintessential field, Eur. Phys. J. C 84 (2024) 574.
\bt{id3} Chen-Kai Qiao and Mi Zhou, Gravitational Lensing of Schwarzschild and Charged Black Holes Immersed in Perfect Fluid Dark Matter Halo, JCAP 12 (2023) 005. 
\bt{ikr1} E. L. B. Junior et al., Gravitational lensing of a Schwarzschild-like black hole in Kalb-Ramond gravity, Phys. Rev. D 110 (2024) 024077.
\bibitem{akiyamal1} K. Akiyama et al., First M87 Event Horizon Telescope Results. I. The Shadow of the Supermassive Black Hole, Astrophys. J. 875 (2019) L1.
\bibitem{akiyamal12} Kazunori Akiyama et al. First Sagittarius $A^*$ Event Horizon Telescope Results. I. The Shadow of the Supermassive Black Hole in the Center of the Milky Way. Astrophys. J. Lett., 930(2):L12, 2022.
\bt{sgra} S. Gillessen et al., AN UPDATE ON MONITORING STELLAR ORBITS IN THE GALACTIC CENTER, Astrophys. J. 837 (2017) 30.
\bibitem{qpo40} M. E. Beer and P. Podsiadlowski, The quiescent light curve and evolutionary state of gro J1655-40, Mon. Not. Roy. Astron. Soc. 331 (2002) 351.
\bibitem{qpo401} S. E. Motta et al., Precise mass and spin measurements for a stellar-mass black hole through X-ray timing: the case of GRO J1655-40, Mon. Not. Roy. Astron. Soc. 437 no. 3, (2014) 2554-2565. 
\bibitem{qpo564} J. A. Orosz et al., An Improved Dynamical Model for the Microquasar XTE J1550-564, Astrophys. J. 730 (2011) 75..
\bibitem{resonance} M.A. Abramowicz et al., Non-linear resonance in nearly geodesic motion in low-mass X-ray binaries, Publ. Astron. Soc. Jpn. 55 (2023) 466.
\bibitem{resonance1} J. Horak and V. Karas, Twin-peak quasiperiodic oscillations as an internal resonance, Astron. Astrophys. 451, 377 (2006).
\bibitem{forced} I. Banerjee,Testing black holes in non-linear electrodynamics from the observed quasi-periodic oscillations, JCAP 08 (2022) 034.
\bibitem{sanjar} S. Shaymatov et al., Charged particle and epicyclic motions around $4D$ Einstein-Gauss-Bonnet black hole immersed in an external magnetic field, Phys. Dark Universe 30 (2020) 100648.
\bibitem{vrba} Z. Stuchlík and J. Vrba, Epicyclic orbits in the field of Einstein-Dirac-Maxwell traversable wormholes applied to the quasiperiodic oscillations observed in microquasars and active galactic nuclei, Eur. Phys. J. Plus 136 (2021) 1127.
\bibitem{QPO1} L. Stella and M. Vietri, kHz Quasi Periodic Oscillations in Low Mass X-ray Binaries as Probes of General Relativity in the Strong Field Regime, Phys. Rev. Lett. 82 (1999) 17-20 (1999).
\bibitem{QPO2} L. Stella and M. Vietri, Lense-Thirring Precession and QPOs in Low Mass X-Ray Binaries, Astrophys. J. Lett. 492 (1998) L59.
\bibitem{QPO3}C. Bambi et al., Testing the no-hair theorem with the continuum-fitting and the iron line methods: a short review, Class. Quant. Grav. 33 (2016) 064001.
\bt{QPO4} C. Bambi, Testing the Kerr-nature of stellar-mass black hole candidates by combining the continuum-fitting method and the power estimate of transient ballistic jets, Phys. Rev. D 85 (2012) 043002
\bibitem{QPO5} M. Tarnopolski and V. Marchenko, A Comprehensive Power Spectral Density Analysis of Astronomical Time Series. II. The Swift/BAT Long Gamma-Ray Bursts, Astrophys. J. 911, 20 (2021).
\bibitem{QPO6} M. Kolos et al., Quasi-harmonic oscillatory motion of charged particles around a Schwarzschild black hole immersed in an uniform magnetic field, Class. Quantum Gravity 32 (2015) 165009.
\bibitem{QPO9} J. Rayimbaev et al., Quasiperiodic Oscillations, Quasinormal Modes and Shadows of Bardeen-Kiselev Black Holes, Phys.Dark Universe 35, 100930 (2022).
\bibitem{QPO10} K. Jusufi et al., Equatorial and polar quasinormal modes and quasiperiodic oscillations of quantum deformed Kerr black hole, Universe 8(4) (2022) 210.
\bt{QPO11} T. Xamidov et al., Probing the Schwarzschild black hole immersed in a dark matter halo through astrophysical tests, Eur. Phys. J. C 85 (2025) 1193.
\bt{QPO12} M. Alloqulov et al., Epicyclic oscillations and accretion disk around a special Buchdahl-inspired spacetime, JHEAP48 (2025) 100424.
\bt{QPO13} M. Guo et al., Parameter constraints on a black hole with Minkowski core through quasiperiodic oscillations, Eur. Phy. J. C 85 (2025) 95.
\bibitem{wein} S. Weinberg, Gravitation, and Cosmology: Principles and Applications of the General Theory of Relativity (New York:Wiley, 1972).

\end{thebibliography}
\end{document}